\documentclass[doublecol]{epl2}
\usepackage{color,soul}

\def\ket#1{\mathinner{|{#1}\rangle}}

\title{Conversion between electromagnetically induced transparency and absorption in a three-level lambda system}
\shorttitle{Conversion between electromagnetically induced transparency and absorption}

\author{Sapam Ranjita Chanu
 \and Kanhaiya Pandey\footnote{Presently at Centre for Quantum Technologies, National University of Singapore.}
 \and Vasant Natarajan\thanks{E-mail: \email{vasant@physics.iisc.ernet.in}}}
\shortauthor{S. R. Chanu \etal}

\institute{Department of Physics, Indian Institute of
Science, Bangalore 560\,012, India}

\pacs{42.50.Gy}{Effects of atomic coherence on propagation,
absorption, and amplification of light; electromagnetically
induced transparency and absorption}
\pacs{32.80.Qk}{Coherent control of atomic interactions with photons}
\pacs{32.80.Xx}{Level crossing and optical pumping}

\abstract{
We show that it is possible to change from a {\it subnatural} electromagnetically induced transparency (EIT) feature to a {\it subnatural} electromagnetically induced absorption (EIA) feature in a (degenerate) three-level $\Lambda$ system. The change is effected by turning on a second control beam counter-propagating with respect to the first beam. We observe this change in the $D_2$ line of Rb in a room-temperature vapor cell. The observations are supported by density-matrix analysis of the complete sublevel structure including the effect of Doppler averaging, but can be understood qualitatively as arising due to the formation of $N$-type systems with the two control beams. Since many of the applications of EIT and EIA rely on the anomalous dispersion near the resonances, this introduces a new ability to control the sign of the dispersion.
}

\begin{document}
\bibliographystyle{eplbib}
\maketitle

Electromagnetically induced transparency (EIT) is an important phenomenon in three-level systems \cite{FIM05}. It has wide-ranging applications such as lasing without inversion \cite{AGA91}, nonlinear optics \cite{HFI90}, suppression of spontaneous emission \cite{GZM91}, high-resolution spectroscopy \cite{RAN02,KPW05}, enhancement of magneto-optic effects \cite{SLB00} (for precision measurements), slowing of light \cite{HHD99} (for use in quantum-information processing), and white-light cavities \cite{WUX08} (for use in gravitational wave detection). The effect arises when a strong control laser on one transition is used to modify the absorption properties of a weak probe laser on a second transition. The control laser creates new {\it dressed states} \cite{COR77} of the combined atom-photon system due to the AC Stark shift, and it is a combination of this energy shift and the interference between the dressed states \cite{LIX95} that gives rise to the EIT phenomenon. In addition, the control laser can cause population redistribution due to optical pumping, which further affects the absorption of the probe. EIT is particularly important in lambda ($\Lambda$) systems, where the presence of two ground levels allows the existence of a dark state \cite{FIM05}. Thus, probe absorption is exactly zero at line center, so that the width of the EIT resonance is {\it subnatural} when the control laser has a sufficiently low power. Even at high powers, the width of the EIT resonance {\it in hot vapor} remains subnatural because of a surprising narrowing effect due to Doppler averaging \cite{KPW05,IKN08}.

The phenomenon of electromagnetically induced absorption (EIA) \cite{LBA99,GWR04,BMW09,PKN11}, though related to EIT, is somewhat less studied; in fact most of the previous work has been in {\it four-level systems}. In EIA, the probe laser shows {\it enhanced absorption} in the presence of the control beam. In this work, we demonstrate that it is possible to convert from EIT to EIA in a three-level $\Lambda$ system simply by turning on a second control beam which is counter-propagating with respect to the first. All the other aspects of the phenomenon remain the same, just the narrow {\it subnatural} resonance changes from a transparency dip to an absorption peak. Many of the applications of EIT rely on the ability to control the {\it size} of the anomalous dispersion near an atomic resonance, the ability to switch from EIT to EIA provides a new technique to additionally control the {\it sign} of the dispersion. For example, this technique can be used to switch from subluminal to superluminal propagation of a probe pulse, as has been proposed in four-level systems \cite{CJG07}. From enhanced transparency to enhanced absorption, this provides an unprecedented degree of ``coherent control'' over the properties of a medium.

We have been careful to use orthogonal polarizations for the two control beams so that they do not form a standing wave. The standing-wave geometry, and the resulting spatial modulation of the absorption and refractive index, has been widely used in previous studies. For example, the formation of an electromagnetically induced grating has been used for all-optical switching of the probe pulse \cite{BRX05}. or to create stationary light pulses \cite{BZL03,LLP09,WAL10}. The periodic modulation has also been explored to create tunable photonic bandgap structures \cite{ARL06,KWD08,GZB10}.

\begin{figure}
\centerline{\resizebox{0.95\columnwidth}{!}{\includegraphics{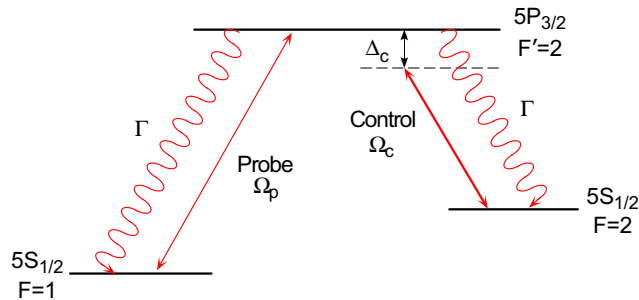}}}
\caption{(Color online) Partial energy-level diagram in Rb showing the transitions of the $D_2$ line used to form the $\Lambda$ system.}
 \label{levels}
\end{figure}

We have performed these experiments in a room-temperature Rb vapor cell using the $5S_{1/2} \rightarrow 5P_{3/2}$ transition, which is the well-known $D_2$ line at 780 nm. The observations are supported by a density-matrix calculation of the complete sublevel structure. The relevant energy levels forming the $\Lambda$ system are shown in Fig.\ \ref{levels}. The two lower levels are the $F=1$ and $F=2$ hyperfine levels of the ground state. The common upper level is the $F=2$ level of the excited state. The control laser is on the $F=2 \leftrightarrow F'=2$ transition, while the probe laser is on the  $F=1 \leftrightarrow F'=2$ transition. The powers in the two lasers are characterized by the respective Rabi frequencies $\Omega$. The decay rate to the ground levels is $\Gamma$, which is $2\pi \times 6$~MHz for these transitions.

\begin{figure}
\centering{\resizebox{0.90\columnwidth}{!}{\includegraphics{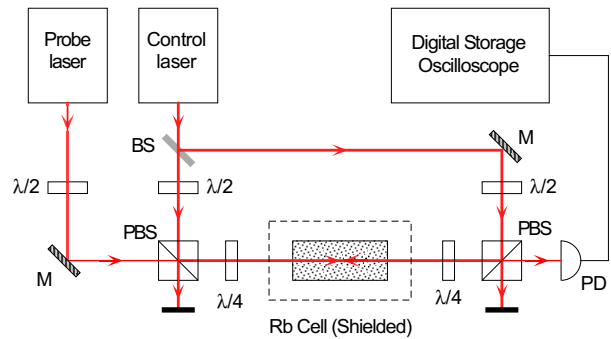}}}
\caption{(Color online) Schematic of the experiment. Figure key: BS -- beamplitter, $\lambda/2$ -- halfwave retardation plate, $\lambda/4$ -- quarterwave retardation plate, M -- mirror, PBS -- polarizing beamsplitter cube, PD -- photodiode.}
 \label{schema}
\end{figure}

The experimental schematic is shown in Fig.\ \ref{schema}. The probe and control beams are derived from two home-built grating-stabilized diode laser systems \cite{BRW01}. The instantaneous linewidth of the lasers after stabilization is about 1~MHz. The beams are elliptic and
have $1/e^2$ diameter of 2~mm $\times$ 3~mm. The probe beam (P) and the co-propagating control beam (CP) have orthogonal polarizations. This allows us to mix and separate them using polarizing beam splitter cubes (PBS's), and detect only the probe beam. The $\lambda/4$ retardation plate ensures that the two beams have orthogonal circular polarizations ($\sigma^-$ for P and $\sigma^+$ for CP) in the experimental cell. The counter-propagating control beam (CO) also passes through a $\lambda/4$ retardation plate, and therefore has the same circular polarization as the probe beam ($\sigma^-$). The experimental cell is a cylindrical vapor cell of dimensions 25-mm diameter $\times$ 50-mm length. The cell
has a three layer magnetic shield that reduces the stray
fields to below 0.1~mG.

The spectra were obtained with the probe laser locked to the $F=2 \leftrightarrow F'=1$ hyperfine transition, while the control laser was scanned around the $F=2 \leftrightarrow F'=2$ transition. Using a locked probe beam and a scanning control beam helps to overcome the first-order Doppler effect \cite{DAN05,IFN09} (this is different from usual EIT experiments where the control beam is fixed and the probe beam is scanned). In effect, the locked probe beam addresses only the zero-velocity atoms, and its absorption remains flat (Doppler free) until it is modified by the control beam. The probe beam was locked using fm modulation spectroscopy in a separate saturated-absorption spectroscopy cell.

\begin{figure}
\centerline{\resizebox{0.9\columnwidth}{!}{\includegraphics{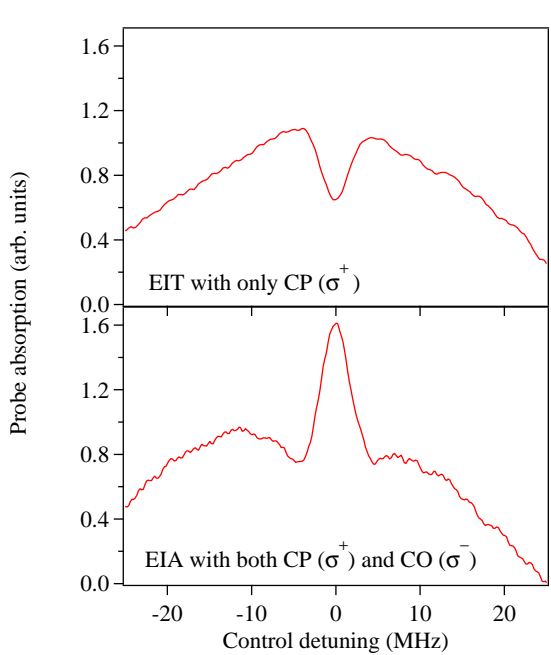}}}
\caption{(Color online) EIT to EIA conversion. Change of lineshape from enhanced probe transmission (EIT) with only the co-propagating control beam to enhanced probe absorption (EIA) with both control beams. The two control beams have orthogonal circular polarizations and powers of 0.9 mW (CP) and 1.5 mW (CO).}
 \label{eiteia}
\end{figure}

The observations are shown in Fig.\ \ref{eiteia}. The probe-absorption spectrum on the top is the typical EIT lineshape with a narrow dip at line center obtained with just the co-propagating control beam on. This geometry with the probe and control beams propagating co-linearly is the standard geometry for observing EIT in a $\Lambda$ system \cite{LIX95}. Not surprisingly, the EIT dip disappears if only the counter-propagating control beam is present instead. The power in the control beam for the spectrum shown is 0.9~mW. If we assume that this power is spread uniformly over the beam size and take into account the $\sim10$\% loss at the cell entrance window, it corresponds to a Rabi frequency of 12~MHz. This means that the dressed states created by the control beam will be located at $\pm 6$~MHz, therefore one does not expect the EIT dip to be {\it subnatural} (less than 6~MHz). However, the observed dip is only 4.4~MHz wide ($0.7 \, \Gamma$). As explained earlier, this is due to the effect of Doppler averaging in room-temperature vapor \cite{IKN08}; the detuned resonance seen by moving atoms fills in the transparency window and makes it narrower. In addition, one expects the spectrum away from line center to be flat in our technique of scanning only the control beam, but there is a broad (30-MHz wide) enhanced absorption peak away from line center. This is due to {\it optical pumping} by the strong control laser, which transfers population into the $F=1$ level and thereby increases probe absorption \cite{IKN08}. The main result of our experiment is seen in the lower spectrum when the counter-propagating control beam (with a power of 1.5~mW) is also turned on. The EIT dip gets inverted into an EIA peak while the width of the resonance remains subnatural at $\sim 4.7$~MHz.

Before trying to understand these line shapes theoretically, we note that the EIA feature that we observe is qualitatively different from the narrow enhanced-absorption feature seen in Ref.\ \cite{LBA99}, where the magnetic sublevels of a transition in the $D_2$ line of Rb were used to form a multilevel system. The earlier work is akin to the phenomenon of coherent population trapping (CPT) with two phase-coherent beams driving the atoms into a bright superposition state. In other words, it is a two-field excitation of the atomic system. The narrow feature (with width of about 100~kHz) is seen when the difference frequency between the two beams is equal to the splitting of the ground levels, so that the ground levels are resonantly coupled to the excited level (the two-photon Raman resonance condition). This feature is {\em not subnatural} because the relevant natural linewidth is the width of the transition between the ground levels, which is sub-Hz because it is electric-dipole forbidden. In that work, the Raman resonance condition corresponded to zero difference frequency between the two beams (because of the use of degenerate sublevels), and the phase coherence was achieved by using a single laser to generate both the beams \cite{LBA99}. By contrast, our EIA feature is created entirely by the control laser, and the phase-independent probe laser only plays the role of measuring this modification.

\begin{figure}
\centering{\resizebox{0.85\columnwidth}{!}{\includegraphics{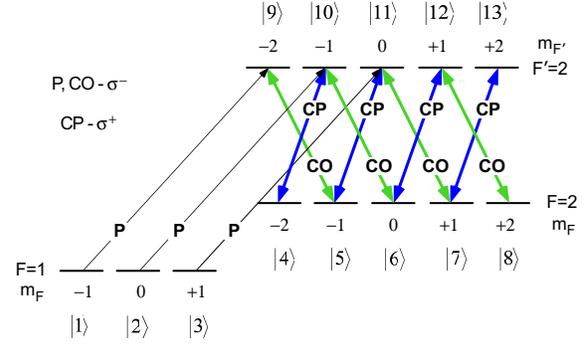}}}
\caption{(Color online) Complete sublevel structure showing the levels coupled by P ($\sigma^-$), CP ($\sigma^+$), and CO ($\sigma^-$). Each sublevel is labeled as $\ket{i}$, with $i$ ranging from 1 to 13.}
 \label{sublevel}
\end{figure}

We have done a density matrix analysis of the complete sublevel structure shown in Fig.\ \ref{sublevel}. The analysis is similar to what we have in our earlier work, in Ref.\ \cite{IKN08} for example. The probe beam and the counter-propagating control beam are $\sigma^-$ polarized, and couple sublevels with the selection rule $\Delta m = -1$. The co-propagating control beam is $\sigma^+$ polarized and couples sublevels with the selection rule $\Delta m = +1$. In the density-matrix approach, probe absorption is proportional to Im($\rho_{i,j}$), where $\ket{i}$ and $\ket{j}$ are the two levels coupled by the probe laser.

\begin{figure}
\centering{\resizebox{0.9\columnwidth}{!}{\includegraphics{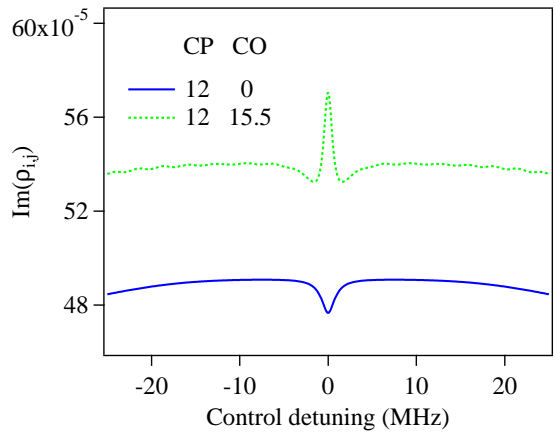}}}
\caption{(Color online) Calculated probe absorption plotted as Im($\rho_{i,j}$) for the two cases. The respective Rabi frequencies (in MHz) are listed.}
 \label{calc}
\end{figure}

When only the CP beam is on, the figure shows that there are two probe couplings, starting from levels $\ket{2}$ and $\ket{3}$, respectively. They both form $\Lambda$-type systems with the control beam: with the levels $\ket{2} \rightarrow \ket{10} \leftrightarrow \ket{4}$ and with the levels $\ket{3} \rightarrow \ket{11} \leftrightarrow \ket{5}$. The total probe absorption is the sum of Im($\rho_{2,10}$) and Im($\rho_{3,11}$), with the coupling-beam strengths weighted by the respective Clebsch-Gordan coefficients. The calculated spectrum {\it after taking into account Doppler averaging} is shown in Fig.\ \ref{calc}.

Now consider what happens when both control beams are on.
Fig.\ \ref{sublevel} shows that there are three couplings due to the probe laser, starting from $\ket{1}$, $\ket{2}$, and $\ket{3}$, respectively. The first forms a double $N$-type system with the levels $\ket{1} \rightarrow \ket{9} \leftrightarrow \ket{5} \leftrightarrow \ket{11} \leftrightarrow \ket{7} \leftrightarrow \ket{13}$. The second forms two systems, a $\Lambda$-type system with the levels $\ket{2} \rightarrow \ket{10} \leftrightarrow \ket{4}$ and an $M$-type system with the levels $\ket{2} \rightarrow \ket{10} \leftrightarrow \ket{6} \leftrightarrow \ket{12} \leftrightarrow \ket{8}$. The third probe forms two $N$-type systems, with the levels $\ket{3} \rightarrow \ket{11} \leftrightarrow \ket{7} \leftrightarrow \ket{13}$ and with the levels $\ket{3} \rightarrow \ket{11} \leftrightarrow \ket{5} \leftrightarrow \ket{9}$. Probe absorption is now proportional to the sum of Im($\rho_{1,9}$), Im($\rho_{2,10}$) and Im($\rho_{3,11}$). As before, the coupling-beam strengths are weighted by the respective Clebsch-Gordan coefficients. The calculated spectrum after taking into account Doppler averaging is shown in Fig.\ \ref{calc}.

The time evolution for the calculated spectra is stopped at 30~$\mu$s, by which time all the transients have died down. The ground-state sublevels are assumed to decohere at a rate of 1~MHz to account for the finite linewidth of the control laser, as has been shown in previous work both theoretically and experimentally \cite{GXL03}. The Rabi frequencies of the two control beams are taken to be 12 MHz and 15.5 MHz respectively, which, as mentioned before, corresponds to the experimental powers if we assume the intensity to be uniformly distributed over the beam size and take into account the 10\% loss at the cell window. There are no other adjustable parameters. The calculations reproduce the main features of the observed spectra, namely EIT with one control beam and EIA with both control beams.

\begin{figure}
\centering{\resizebox{0.9\columnwidth}{!}{\includegraphics{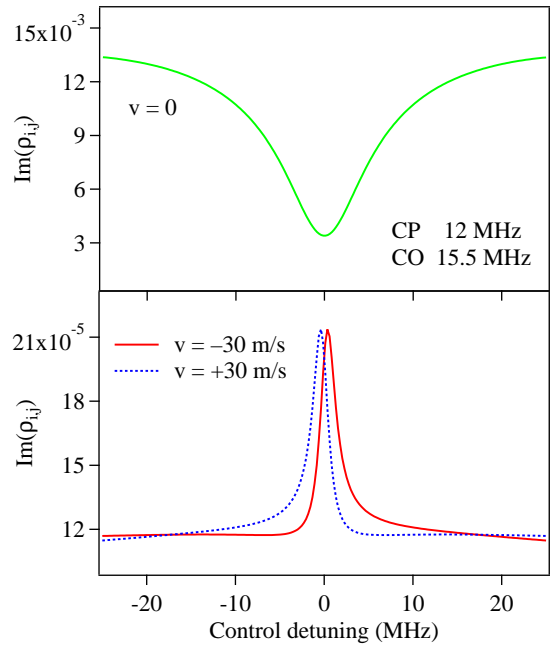}}}
\caption{(Color online) Effect of velocity on probe absorption with both control beams on. Im($\rho_{i,j}$) is shown for zero velocity atoms (upper part) and for atoms moving with $v = \pm 30$~m/s (lower part), defined with respect to the direction of the probe beam. Note the change in scale of the $y$-axis for the two graphs.}
 \label{dopp}
\end{figure}

With only the CP control beam, the EIT feature is expected because it a $\Lambda$-type system \cite{IKN08}. With both control beams, the EIA feature arises mainly due to the formation of $N$-type systems and the effect of Doppler averaging in room temperature vapor. Further justification for the importance of forming an $N$-type system is given below when we discuss the alternate $\Lambda$ system. But the role of thermal averaging can be understood by referring to Fig.\ \ref{dopp}. For stationary atoms ($v=0$), the calculated absorption remains a broad EIT dip, as seen in the upper curve. However, this gets converted to an EIA peak for moving atoms, which see the probe and two control beams as being detuned from resonance. This is seen in the lower part of the figure, calculated for atoms moving with a velocity of $\pm 30$~m/s.The change is accompanied by a large change in the scale of the absorption. In other words, the large EIT dip for stationary atoms gets converted to a small EIA peak for moving atoms. The average over all the velocity classes results in the final EIA peak shown in Fig.\ \ref{calc}.

While the calculated spectra reproduce the main observations, the calculated linewidths for both the EIT and EIA features are smaller than the observed values. This is probably because of the 1-MHz linewidth of the probe laser, which is not included in the calculation. The resonance can also be broadened if there is a small misalignment angle between the beams \cite{CAT04}.

\begin{figure}
(a)\centering{\resizebox{0.85\columnwidth}{!}{\includegraphics{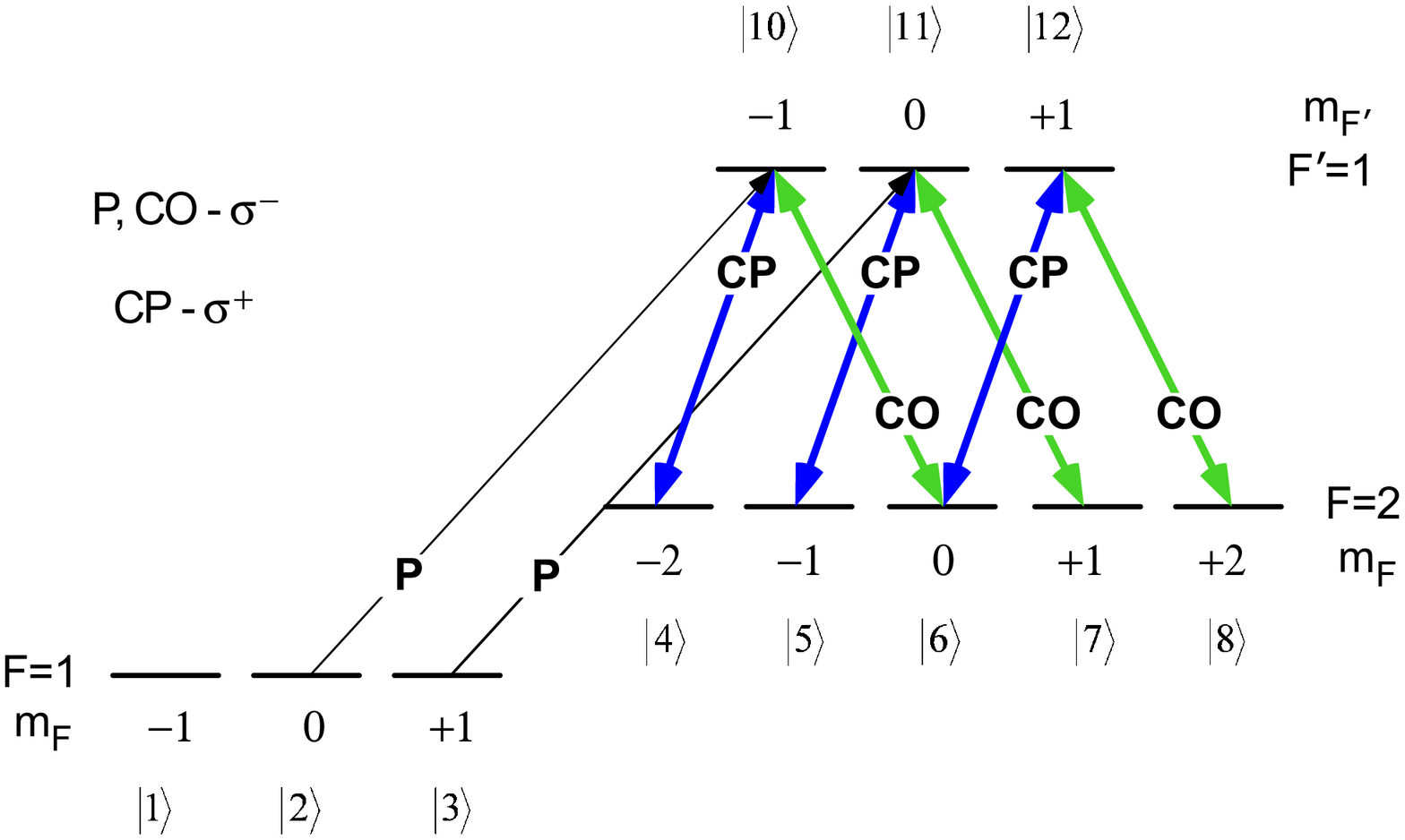}}}
(b)\centering{\resizebox{0.9\columnwidth}{!}{\includegraphics{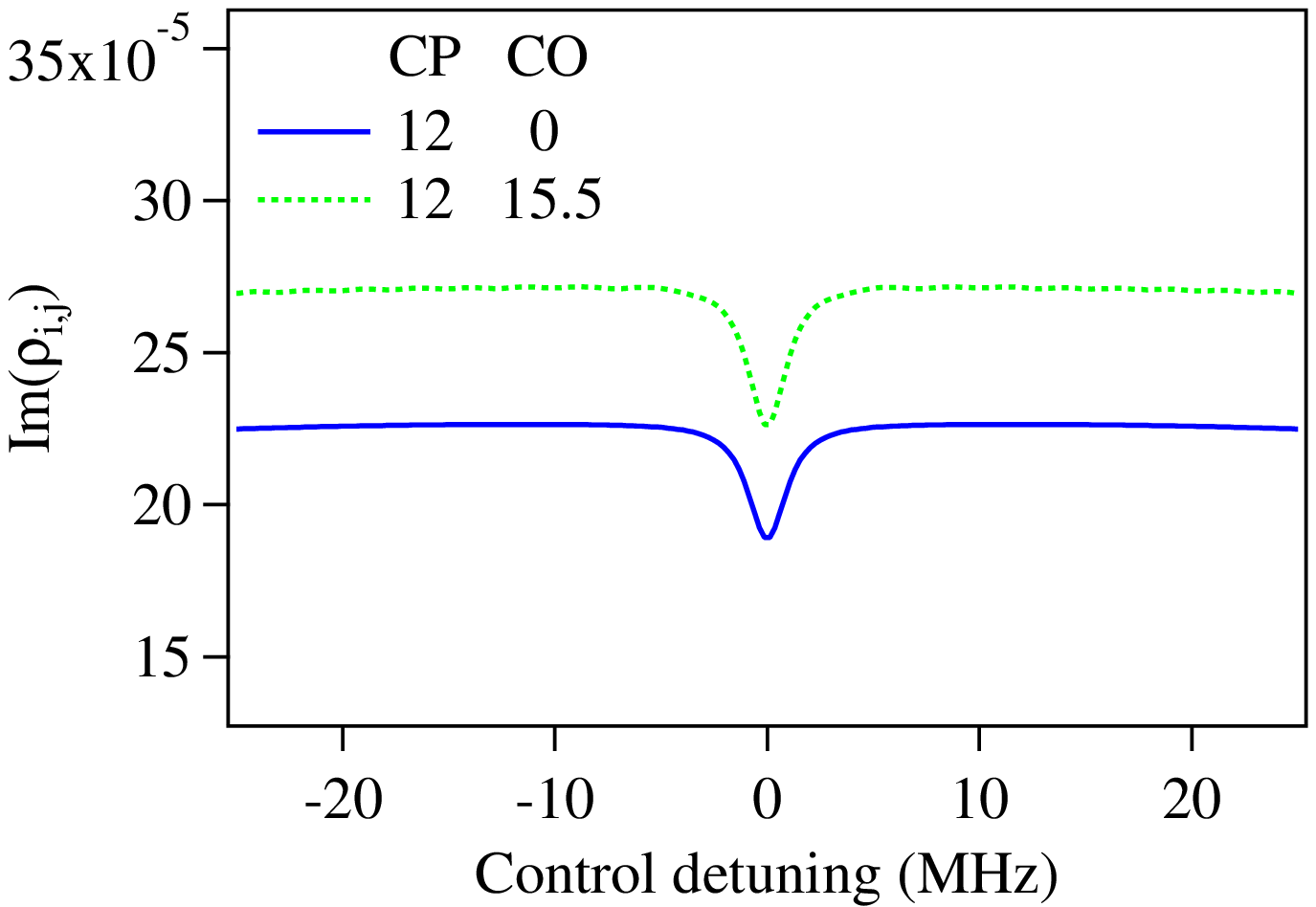}}}
\caption{(Color online) (a) Complete sublevel structure for the alternate $\Lambda$ system formed using the $F=1 \rightarrow F'=1 \rightarrow F'=2$ levels. (b) Calculated probe-absorption, plotted as Im($\rho_{i,j}$), for the two cases of just CP on and CP and CO on. The respective Rabi frequencies (in MHz) are listed. Both cases show only EIT as observed experimentally.}
 \label{sublevel2}
\end{figure}

Our experiments above were done using the upper $F'=2$ level to form the $\Lambda$ system. But an alternate $\Lambda$ system can be formed using the upper $F'=1$ level. Although this shows EIT with just the co-propagating control beam \cite{IKN08}, it does not convert to EIA when the counter-propagating control beam is turned on. This is because the upper level has a smaller number of sublevels, which prevents the formation of any $N$-type system. As seen in Fig.\ \ref{sublevel2}(a), the removal of sublevels $\ket{9}$ and $\ket{13}$ from Fig.\ \ref{sublevel} results in the formation of three $\Lambda$-type systems and just one $M$-type system (with the levels $\ket{2} \rightarrow \ket{10} \leftrightarrow \ket{6} \leftrightarrow \ket{12} \leftrightarrow \ket{8}$). The non-conversion from EIT to EIA is borne out by the density-matrix calculations, shown in Fig.\ \ref{sublevel2}(b). In a recent study \cite{CSB11}, we have shown that EIA can also be observed in a {\em degenerate  two-level transition} with just one control laser. There again, the transition was from $F$ to $F+1$, so that the larger number of magnetic sublevels in the upper level allowed the formation of an $N$-type system. Therefore, it seems that the presence of $N$-type systems is crucial for getting EIA \cite{BMW09}.

In conclusion, we have shown that it is possible to convert from enhanced probe transmission to enhanced probe absorption in a degenerate three-level $\Lambda$ system. We get the usual subnatural EIT feature when one (co-propagating) control beam is on, which transforms to EIA when a second (counter-propagating) control beam is turned on. The observations are explained by density-matrix analysis of the complete sublevel structure after accounting for the effects of the thermal velocity distribution in room temperature vapor. But, qualitatively, the change can be understood to arise due to the formation of $N$-type systems. This ability to invert the peak in a straightforward way presents a new technique to control the sign of the dispersion near an atomic transition. It should have a significant impact in many applications of EIT, such as the ability to switch from sub to superluminal light propagation \cite{CJG07}. In addition, we have observed that detuning the control beams from resonance makes the lineshape dispersive, which can be used for enhancement of nonlinear effects \cite{NGL05}.

\begin{acknowledgments}
This work was supported by the Department of Science and
Technology, Government of India. One of us (K.P.) acknowledges financial support from the Council of Scientific and Industrial Research, India.
\end{acknowledgments}


\end{document}